\documentclass[10pt,aps,twocolumn,superscriptaddress]{revtex4-1}
\usepackage[centertags]{amsmath}
\usepackage{amsfonts,amssymb, amsthm}
\usepackage{hyperref}
\usepackage{comment}

\usepackage{graphicx}  
\usepackage{amsmath}
\usepackage{amsmath,amssymb}
\usepackage{hyperref}
\usepackage{color}
\usepackage[sort&compress]{natbib}
\usepackage{amsmath,amssymb,MnSymbol,dsfont}

\usepackage{subcaption}
\usepackage{ulem}

\begin{document}

\date{\today}
\title{Anomalous conduction and second sound in the Fermi-Pasta-Ulam-Tsingou chain: wave-turbulence approach }
\author{Francesco {De Vita}}\affiliation{DMMM, Politecnico di Bari, Via Re David 200, Bari, 70125, Italy} 
\author{Giovanni \surname{Dematteis}}\affiliation{Department of Mathematical Sciences, Rensselaer Polytechnic Institute, Troy, NY (US)}
\author{Raffaele \surname{Mazzilli}}\affiliation{Max-Planck-Institut f\"ur Festk\"orperforschung, 70569 Stuttgart, Germany}
\author{Davide \surname{Proment}}\affiliation{School of Mathematics, University of East Anglia, Norwich Research Park, NR4 7TJ Norwich, United Kingdom}
\author{Yuri V. \surname{Lvov}}\affiliation{Department of Mathematical Sciences, Rensselaer Polytechnic Institute, Troy, NY (US)}
\author{Miguel \surname{Onorato}}\affiliation{Dipartimento di Fisica, Universit\`{a} di Torino, Via P. Giuria, 1 - Torino, 10125, Italy}\affiliation{INFN, Sezione di Torino, Via P. Giuria, 1 - Torino, 10125, Italy}

\begin{abstract}
One-dimensional particle chains are fundamental models to explain anomalous thermal conduction in low-dimensional solids like nanotubes and nanowires. In these systems the thermal energy is carried by phonons, i.e. propagating lattice oscillations that interact via nonlinear resonance. The average energy transfer between the phonons is described by the wave kinetic equation (WKE), derived directly from the microscopic dynamics.
Here, we use the spatially nonhomogeneous WKE of the prototypical $\beta-$Fermi-Pasta-Ulam-Tsingou (FPUT) model, equipped with thermostats able to set different temperatures at the two ends. Our main findings are as follows: (i) The anomalous scaling of the conductivity with the system size, in close agreement with the known results from the microscopic dynamics, is due to a nontrivial interplay between high and low wavenumbers. (ii) The high-wavenumber phonons relax to local thermodynamic equilibrium transporting energy diffusively, {\it \`a la Fourier}. (iii) The low-wavenumber phonons are nearly noninteracting and transfer energy ballistically;  this latter phenomenon is the analogous of the second sound emission, observed for example in superfluids.
\end{abstract}

\maketitle




{T}hat heat conduction in three-dimensional macroscopic solids is described effectively by Fourier's law has been known for two centuries~\cite{baron1822theorie}. This law establishes the existence of a size-independent {\it property of the material}, the heat conductivity $\mathcal{K}$, as a finite proportionality constant between the heat flux and its driving {\it thermodynamic force}, the temperature gradient~\cite{kubo2012statistical}. In low-dimensional solids this proportionality may break down and size-dependent conduction effects arise. One-dimensional (1D) particle chains characterized by harmonic potential have a conductivity proportional to the chain length $L$, that is $\mathcal{K}\propto L^{1}$~\cite{rieder1967properties}. The lattice excitations propagate unperturbed as non-interacting wave packets; this implies absence of relaxation. Therefore, energy is transported by advection (or {\it ballistically}~\cite{dhar2016heat}), rather than by diffusion as prescribed by Fourier's law. Intermediate behaviors in which $\mathcal{K}\propto L^{\alpha}$, with $0<\alpha<1$, are observed both in numerical simulations of more realistic particle chains~\cite{lepri2003thermal,donadio2009atomistic} and in experiments involving nearly 1D systems such as carbon nanotubes and silicon nanowires~\cite{chang2008breakdown,chen2010remarkable,shen2010polyethylene,yang2010violation,liu2012anomalous,chang2016experimental}. This type of transport, neither diffusive nor ballistic, is referred to as {\it anomalous} and its investigation, motivated by the relevance of the widespread technological applications, has generated a large body of literature in the last two decades~\cite{lepri2016thermal}.

A minimal model displaying anomalous transport is the Fermi-Pasta-Ulam-Tsingou (FPUT) chain~\cite{fermi1955alamos,gallavotti2007fermi}, in which a cubic ($\alpha-$FPUT) or quartic ($\beta-$FPUT) anharmonic term is added to the harmonic potential. Physically, this term represents the lowest-order nonlinear correction to the linearized harmonic system, in a power-law expansion of the potential around the equilibrium point. 
Weak nonlinearity plays a crucial role in the derivation of the so-called {\it phonon Boltzmann equation}, or {\it wave kinetic equation} (WKE)~\cite{peierls1929,zakharov2012kolmogorov,choi2005joint,nazarenko2011wave,eyink2012kinetic}. The WKE is the statistical closure of the deterministic equations of motion, for the {\it spectral action density} $n(k,t)$, {\it i.e.} the second moment of the random wavefield in Fourier space; $n(k,t)$ is related to the spectral energy density $e(k)$ via $e(k)=\omega(k)n(k)$, where $\omega(k)$ is the linear dispersion relation. The WKE contains the statistical description of the action/energy transfers between the various wave modes. These  transfers are mediated by the {\it collision integral}, which incorporates the nonlinear terms in the WKE as wave-wave resonant interactions between the eigenstates of the harmonic chain, i.e. the linear waves. The time scales of the resonant transfers, the {\it kinetic time scales}, are much longer than the linear time scale of wave propagation, where the scale separation is tied to the smallness of the nonlinearity. For $\beta-$FPUT in the thermodynamic limit (large system), the quartic term in the potential yields a collision integral encoding $4-$wave resonant interactions~\cite{chibbaro2017wave,chibbaro20184,onorato2020straightforward} that exchange energy between quartets of wave modes and, ultimately, lead to relaxation over the kinetic time scales~\cite{lvov2018double}.

A kinetic interpretation of the numerically observed anomalous exponent $\alpha\simeq0.4$~\cite{lepri1998anomalous} in the $\beta-$FPUT system was proposed in~\cite{pereverzev2003fermi} and given rigorous justification in~\cite{aoki2006energy,lukkarinen2008anomalous}. The result is based on an asymptotic approximation of the collision integral for the {\it acoustic modes}, as $k\to0$, and on a subsequent heuristic cut-off applied to the linear-response Kubo integration of the correlation function~\cite{lepri2003thermal}. In~\cite{Dematteis2020}, an analogous asymptotic result was exploited to identify a critical scale $k_c$ determining a separation between two sets of modes: those with $|k|<k_c$ essentially behave as ballistic modes in an harmonic chain, as they lack a sufficient level of interaction; the modes with $|k|>k_c$ are diffusive and relax locally to the expected Fourier profile. In this interpretation, the anomalous exponent is due to how the separation between the two sets scales with the chain length, given by a certain function $k_c(L)$. Rather than from an anomalous type of diffusion (such as fractionary diffusion), the anomalous transport properties thus arise from a coexistence of regular diffusion and advective effects that persist on macroscopic scales due to the properties of the collision integral.

Here, we go beyond the asymptotic result for the acoustic modes and fully exploit the mesoscopic picture, by direct numerical integration in time of the spatially-nonhomogeneous WKE associated with the $\beta-$FPUT model. We show that the first-principle description of the WKE is able to reproduce accurately the anomalous transport properties observed in the direct numerical simulations of the microscopic dynamics~\cite{lepri2016thermal}. We also confirm results previously conjectured on the separation between ballistic modes and diffusive modes~\cite{Dematteis2020}, and characterize the respective transport coefficients. Finally, we exploit the higher-level, and computationally much cheaper picture of the WKE, to investigate  in detail the local relaxation properties of the different modes.

\section*{Model}
To simulate thermal energy transfers through a one-dimensional particle chain we solve the following space-dependent WKE associated with the $\beta-$FPUT model in the thermodynamic limit~\cite{lvov2018double}, for the spectral action density $n_k=n(x,k,t)$~\cite{nazarenko2011wave,deng2021full,ampatzoglou2021derivation}
\begin{equation}
  \label{eqn:wavkin}
  \frac{\partial n_k}{\partial t} + v_k\frac{\partial n_k}{\partial x} = \mathcal{I}_k.
\end{equation}
We denote by $x\in[0,L]$, $k\in[0,2\pi)$ and $t>0$ the (macroscopic) physical space, the Fourier space and the time variables, respectively.
The second term at the LHS of \eqref{eqn:wavkin} represents the advection of $n_k$ due to spatial inhomogeneities, where $\omega_k=2\sin(k/2)$ is the linear dispersion relation and $v_k = d\omega_k/dk = \cos(k/2)$ is the group velocity; the RHS of \eqref{eqn:wavkin} is the $4-$wave collision integral
\begin{equation}
  \label{eqn:coll}
  \begin{aligned}
    \mathcal{I}_k &= 4\pi\int_{0}^{2\pi}|T_{k123}|^2n_{k}n_{k_1}n_{k_2}n_{k_3}\biggl(\frac{1}{n_{k}} + \frac{1}{n_{k_1}} \\ 
    &\quad- \frac{1}{n_{k_2}} - \frac{1}{n_{k_3}}\biggl)\delta(\Delta K)\delta(\Delta \Omega)dk_1dk_2dk_3,
  \end{aligned}
\end{equation}
where the arguments of the Dirac's deltas are defined as $\Delta K = k + k_1 - k_2 - k_3$, $\Delta \Omega = \omega_k + \omega_1 - \omega_2 - \omega_3$, and
$|T_{k123}|^2 = 9\omega_{k}\omega_{1}\omega_{2}\omega_{3}/16$ is the matrix element associated with the $\beta-$FPUT model~\cite{lukkarinen2008anomalous,bustamante2019exact}.
The integration of \eqref{eqn:coll} has only one degree of freedom, since the resonance conditions imposed by the Delta functions constrain the integration to the so-called {\it resonant manifold}, i.e. the subset of possible combinations of $k_1,k_2,k_3$ that are in resonance with mode $k$, representing all the resonant wave quartets.
We provide an explicit expression of this one-dimensional integration in the {\it Materials and Methods} section.
%
%

The resonant interactions contained in the collision integral represent the mechanism responsible for the local ({\it i.e.} at fixed $x$)
relaxation to the equilibrium distribution of $n_k$ which, given the two conserved quantities of \eqref{eqn:wavkin}, is given by the Rayleigh-Jeans (RJ)
solution
\begin{equation}\label{eq:5}
  n^{(RJ)}_k = \frac{T}{\omega_{k} + \mu}\,.
\end{equation}
Here, $T$ plays the role of the temperature of the system and $\mu$ of the chemical potential; these quantities are associated with the conservation
of the harmonic energy and of the action (or number of particles), respectively.  The spatial energy density profile can be 
computed multiplying $n_k$ by $\omega_k$ and integrating in $k$:
\begin{equation}
  e(x,t) = \int_0^{2\pi} \omega_k n(x,k,t)dk. 
\end{equation}

To avoid confusion, we recall that due to the discreteness of the physical space the Fourier space is periodic and therefore the modes in the interval $[\pi,2\pi)$ can be equivalently interpreted as in $[-\pi,0)$. For this reason, hereafter we refer to the modes near $0$ or $2\pi$ as the {\it low wavenumbers} and to the modes near $\pi$ as the {\it high wavenumbers}.
 

\section*{Results}
In what follows, we will discuss results achieved from two types of numerical simulations of the nonhomogeneous wave kinetic equation: case (A) corresponds to the classical problem of a chain in between two thermostats at different temperatures and case (B) corresponds to the free evolution of an initial energy density narrow Gaussian profile in $x$. The latter is the typical experiment used to asses the diffusive (or not) properties of the system. 
\subsection{Anomalous conduction}

\begin{figure*}[ht]
  \centering
  \begin{subfigure}{\columnwidth}
    \includegraphics[width=\columnwidth]{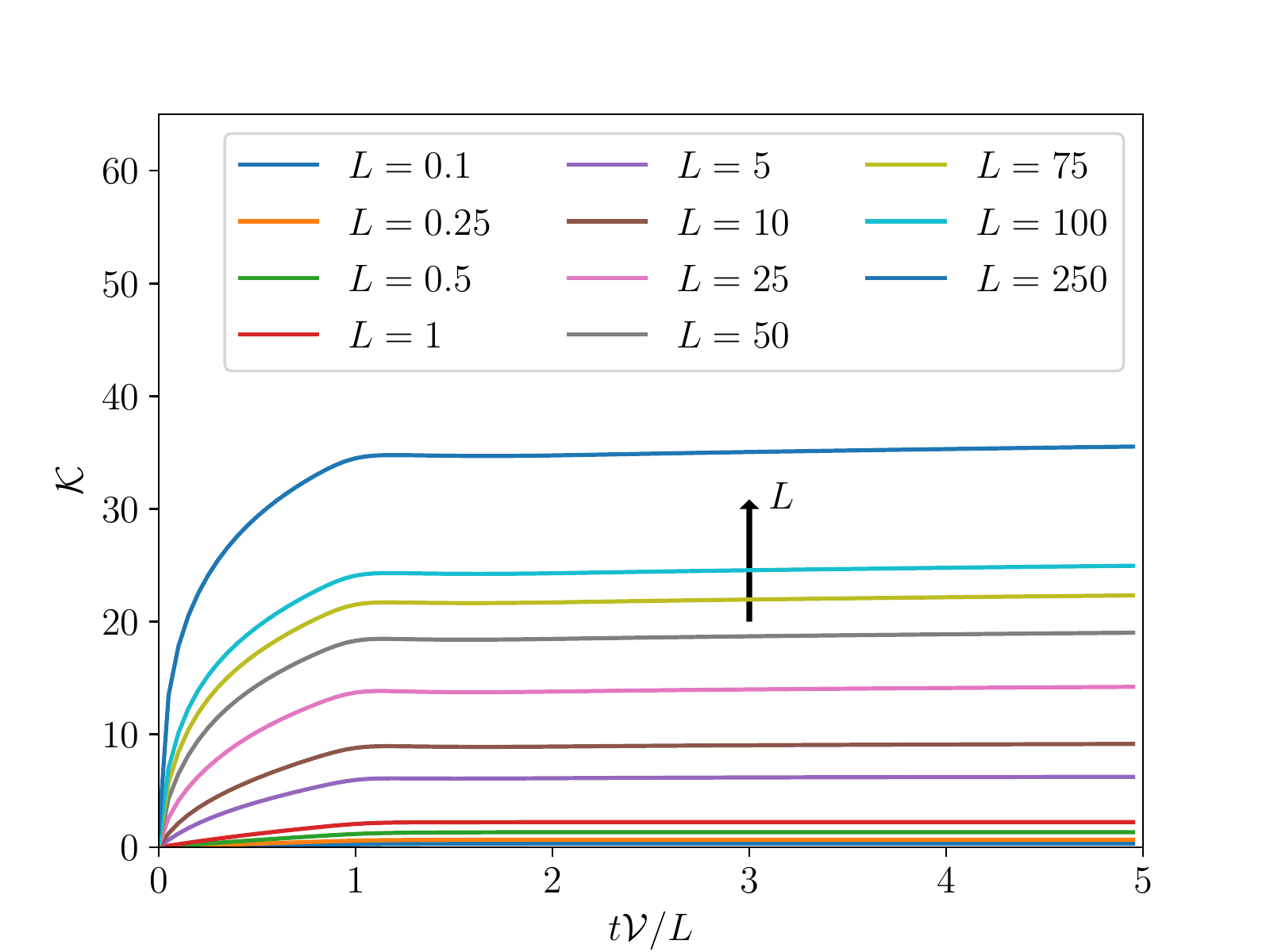}
    \caption{\label{fig:KT}}
  \end{subfigure}
  \begin{subfigure}{\columnwidth}
    \includegraphics[width=\columnwidth]{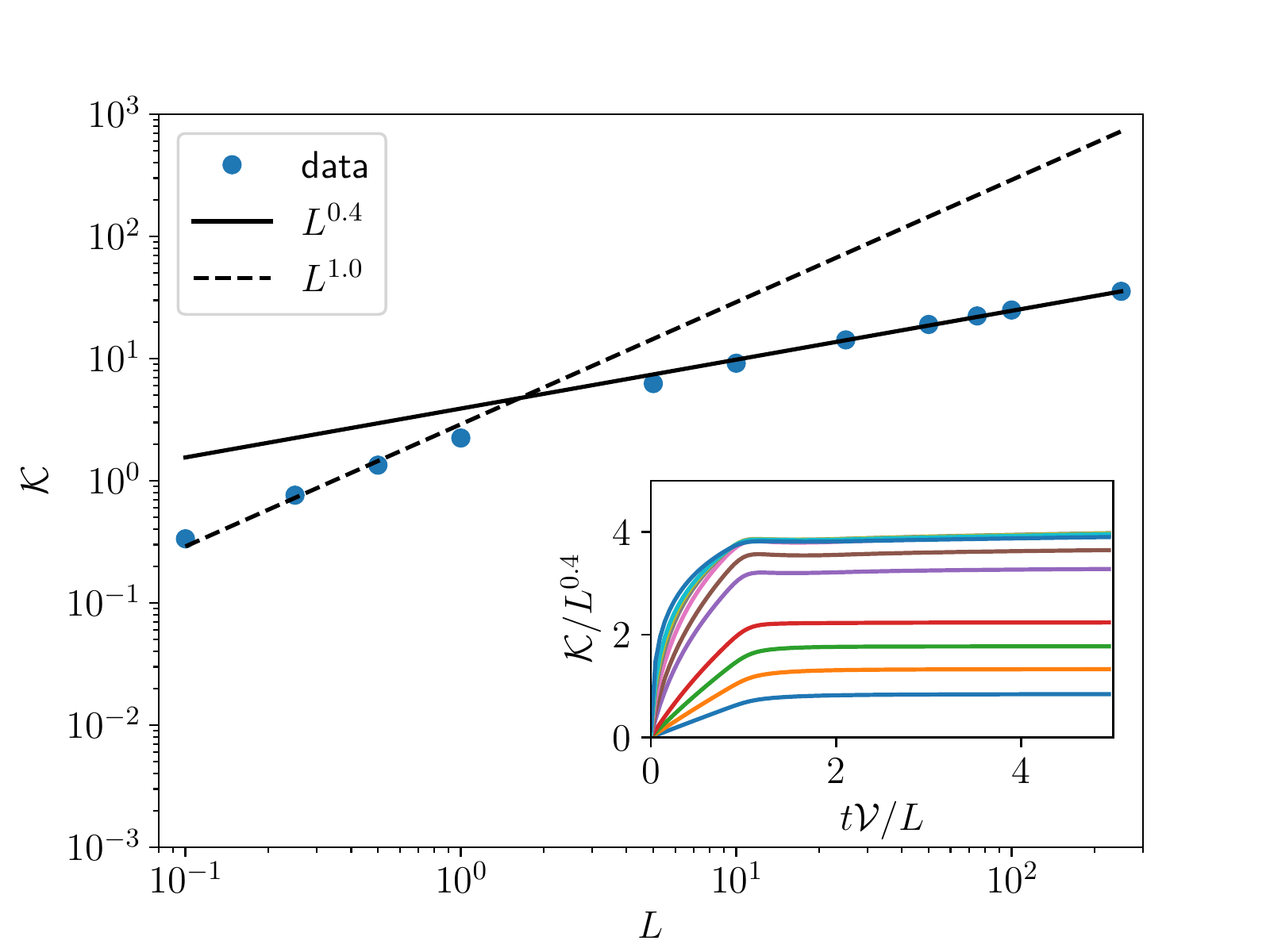}
    \caption{\label{fig:KL}}
  \end{subfigure} 
  \caption{(a) Thermal conductivity $\mathcal{K}$ as function of non-dimensional time for several values of $L$; (b) Steady state thermal conductivity $\mathcal{K}$ 
    as a function of $L$. For small $L$ most of the modes are non-interacting and the ballistic scaling $\mathcal K\propto L^1$ is recovered. For larger $L$, we observe excellent asymptotic agreement with the scaling $\mathcal K\propto L^{0.4}$. The inset of panel (b) reports the thermal conductivity $\mathcal{K}$ in panel (a) normalized by $L^{0.4}$ as a function of non-dimensional time.\label{fig1}}
\end{figure*}

\begin{figure*}[hbt]
  \centering
  \includegraphics[width = \textwidth]{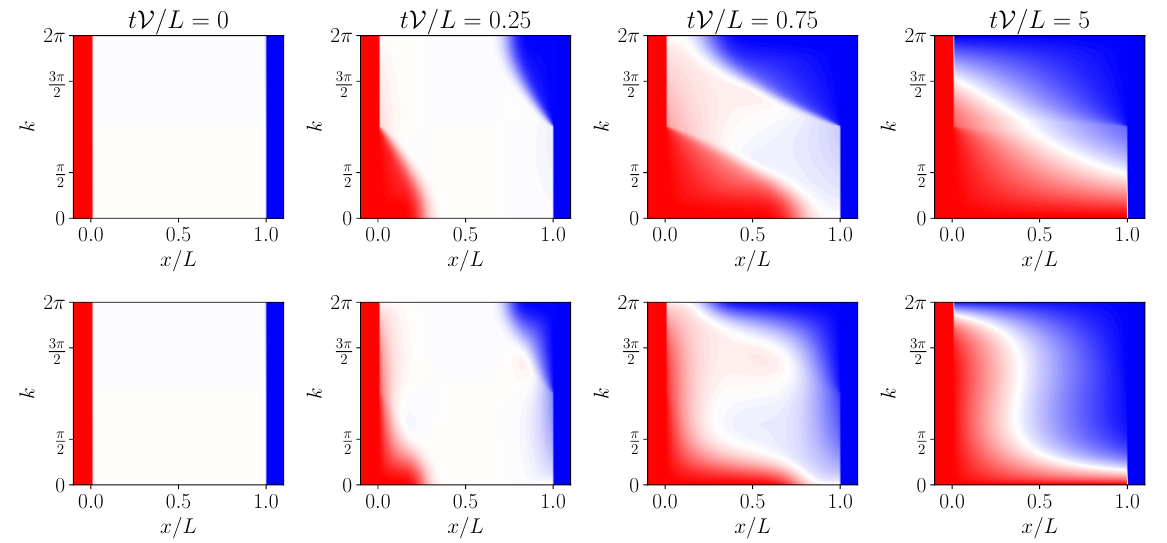}
  \caption{Color map of $T(x,k,t)=(\omega_k+\mu) n(x,k,t)$. Top row: $L=1$; bottom row: $L=10$. Color ranges from $T_2$ (white) to $T_1$ (red). For $t>0$, two fronts start propagating the temperature of the thermostats, perturbing the initial homogeneous state. The modes in $[0,\pi]$ propagate to the right ($v_k>0$) and the modes in $[\pi,2\pi]$ propagate to the left ($v_k<0$). The upper panels depict a predominantly ballistic situation also at the stationary state (right panel), since $L$ is not sufficiently large for most of the modes to interact. For the larger system in the lower panels, once a steady state is reached the diffusive modes (around $k=\pi$) have equipartitioned at fixed $x$ and are
accompanied by a $k-$independent constant gradient
between the two thermostats. On the other
hand, the ballistic modes (around $k=0$ and $k=2\pi$) carry the energy density
of their originating thermostat all the way
to the opposite side without interactions with
other modes. Here we observe that the width of the ballistic region becomes thinner as $L$ increases (see Fig.~\ref{fig:kjmax}).
    \label{fig:Tmap}}
\end{figure*}

To demonstrate numerically anomalous conduction, we consider a domain of size $L$ with two thermostats at its ends at 
different temperature $T_1$ and $T_2$. For normal conduction, at the steady state one expects a linear 
temperature, $T$, profile (Fourier's law) and the conductivity, $\mathcal{K}$, to be independent of the 
size of the domain, $L$. By defining the net spectral energy current as
\begin{equation}
  j(k,x,t) = \omega_k v_k[n(k,x,t) - n(-k,x,t)]/2,
\end{equation}
the conductivity can be computed as
\begin{equation}
  \label{eqn:cond}
  \mathcal{K} = \frac{J L}{\Delta T}
\end{equation}
with 
\begin{equation}
  J = \frac{1}{L}\int_0^L\int_0^{2\pi} j(k,x,t)dkdx,
\end{equation} 
being the spatial average of the integral of $j(k,x,t)$ and $\Delta T = T_1 - T_2$ 
the temperature difference between the two thermostats. The term $\Delta T/L$ represents the mean 
temperature gradient and one can recognize the definition of $\mathcal{K}$ as given by Fourier's law. 
Note that, at the steady state, the energy current is independent of $x$. 

In Fig. \ref{fig:KT}, we report the time history of the conductivity for several values 
of the domain size $L$, keeping fixed $ \Delta T$. The initial ($t=0$) distribution is set to be a RJ distribution at the average  temperature between the two thermostats; subsequently, there is an initial transient during which the energy flux starts growing (and consequently also $\mathcal{K}$ ), until a stationary state is reached.  Note that time is made non-dimensional with the 
reference time $L/\mathcal{V}$, with $\mathcal{V}=v_{k=0}=1$ being the maximal ballistic velocity.
In Fig.~\ref{fig:KL}, instead, we report the measured conductivity as a function of the domain size 
$L$: the results clearly indicate that for small $L$ the stationary value of the conductivity tends to be proportional to $L$, as 
in the purely harmonic system; on the other hand, for $L \to \infty$ the exponent $\alpha$ tends to the 
constant value of $0.4$, consistent with the value  measured in the microscopic simulations \cite{lepri1998anomalous}. In the inset of the figure, the time history of the conductivity 
$\mathcal{K}$ divided by $L^{0.4}$ is drawn to highlight that for large values of $L$ the 
curves overlap.

\begin{figure*}[ht]
  \centering
  \begin{subfigure}{\columnwidth}
    \includegraphics[width = \columnwidth]{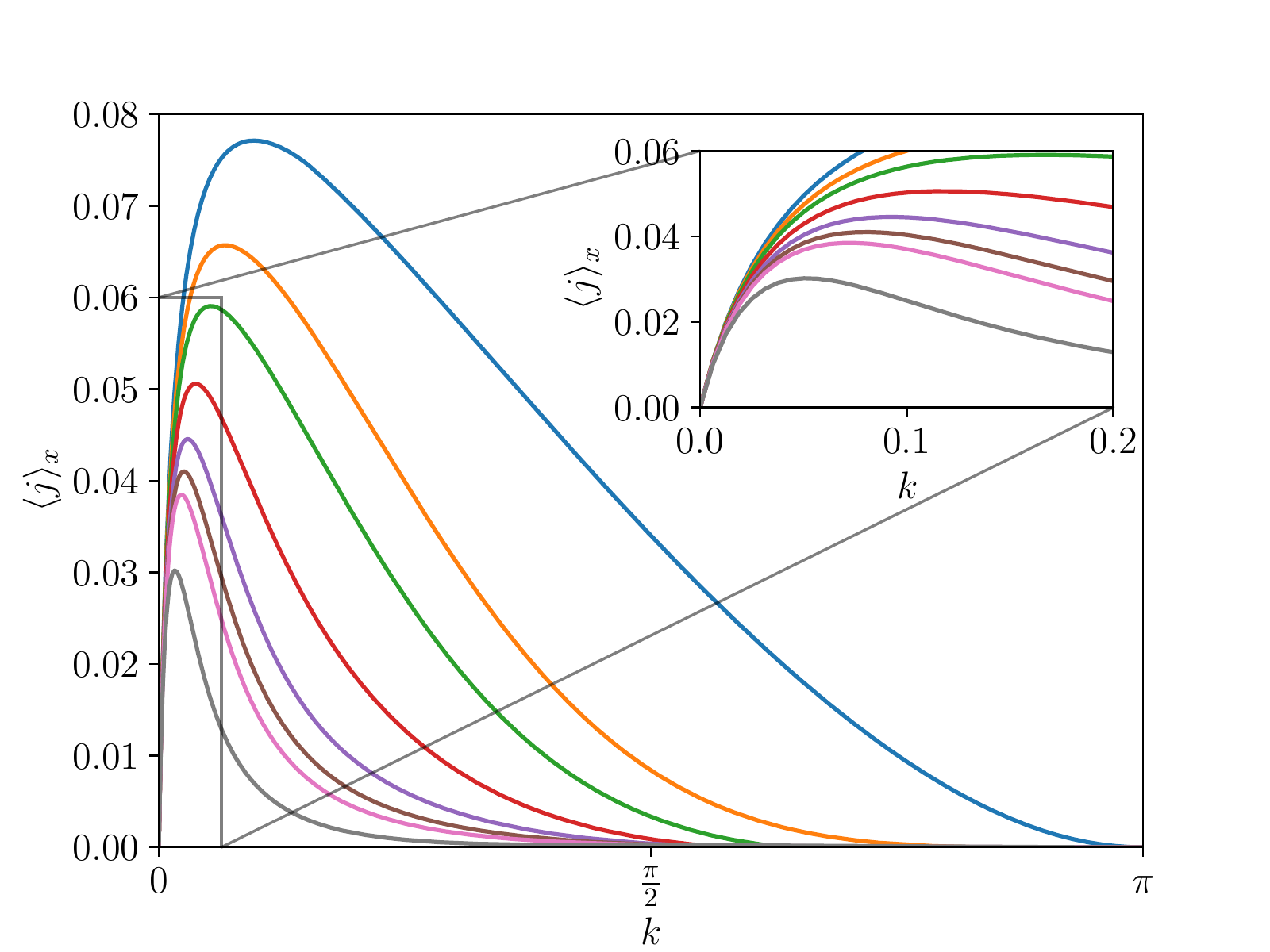}
    \caption{\label{fig:jk}}
  \end{subfigure}
  \begin{subfigure}{\columnwidth}
    \includegraphics[width = \columnwidth]{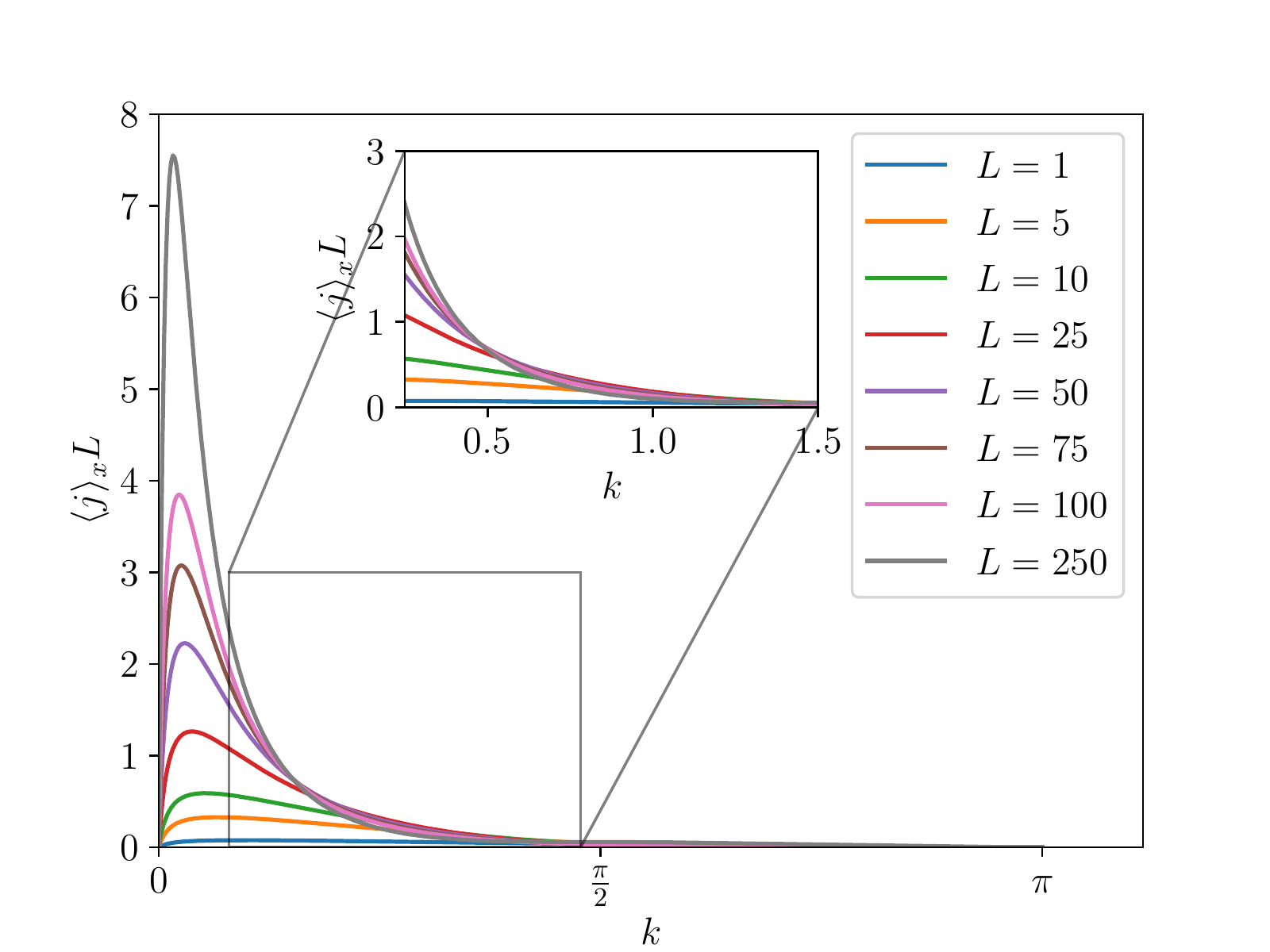}
    \caption{\label{fig:jLk}}
  \end{subfigure} 
  \caption{(a) Energy current $\langle j\rangle_x$ as function of $k$ for different domain size, once a steady state has been reached. While  for the high wavenumbers the energy current contribution decreases as $L$ increases, the low-wavenumber contribution is independent of $L$, in agreement with a ballistic behavior. In the inset, one can appreciate this invariance as $k\to0$.
    (b) The energy current $\langle j\rangle_x$ is multiplied by the size of the domain $L$. Now, the renormalized curves tend to converge onto each other independently of $L$ for the high wavenumbers. This behavior is in agreement with Fourier's law (see \eqref{eqn:cond}) prescribing inverse proportionality between energy current  and system size.
    \label{fig:j}}
\end{figure*}

Via integration of the deterministic microscopic equations, a recent work \cite{Dematteis2020} provided 
evidence that the collision integral $\mathcal{I}_k$ brings the system to local equilibrium down to a 
critical $k_c$ whereas, for lower $k$'s advection is predominant and waves travel in the domain transported 
by the group velocity $v_k$ interacting too weakly in order to relax locally to a RJ spectrum.
Here, we observe this clearly by looking at the evolution of the color map of the temperature in Fig.~\ref{fig:Tmap}, where the temperature spectral density $T(x,k,t)=(\omega_k+\mu)n(x,k,t)$ is defined by inverting \eqref{eq:5}, using the fact that $\mu$ is constant throughout the evolution. Note that for 
$k < \pi$ the velocity $v_k$ is positive, while it is negative for $k > \pi$. The initial condition at 
$t=0$ is a homogeneous field, with $n(x,k,t=0)=n^{(RJ)}_k$ at temperature $(T_1+T_2)/2$. Due to the presence of the thermostats, as 
$t>0$ the waves going to the right start to propagate a hot front from the left thermostat, while 
the waves going to the left start propagating a cold front from the right thermostat. The edge of the 
front propagates at the maximal speed allowed, which is the speed of the acoustic modes 
$v(k\to0^\pm)=\pm1$. For $L = 1$, the energy flows in a ballistic way for almost the entire domain and the 
collision integral is not strong enough to bring the system to local equilibrium. As a result, at large 
times and at any fixed point $x$, the right-going waves are at temperature $T_1$ and the left-going waves 
are at temperature $T_2$, far from local equipartition. For larger system size, $L=10$, instead, advection 
is predominant only in a small range of $k$, with the remaining part of the domain dominated by 
diffusion: in this region, at large time and at any fixed point $x$, the energetic content of the left- 
and right-going waves is equipartitioned, and there is a constant smooth temperature gradient between the 
two thermostats. This is reflected in the profile in $k$ of the spatial integral of the spectral energy 
current $\langle j\rangle_x = \tfrac{1}{L}\int_x j(k,x)dx$, once a steady state is reached, see figure \ref{fig:jk}. For modes with small $k$, $\langle j\rangle_x$ is 
independent of the system size since the behavior is purely ballistic~\cite{rieder1967properties}. For 
modes with large $k$ the energy current flattens as $L$ increases, accompanied by a reduction of the peak 
value.
\begin{figure}
  \centering
  \includegraphics[width=\columnwidth]{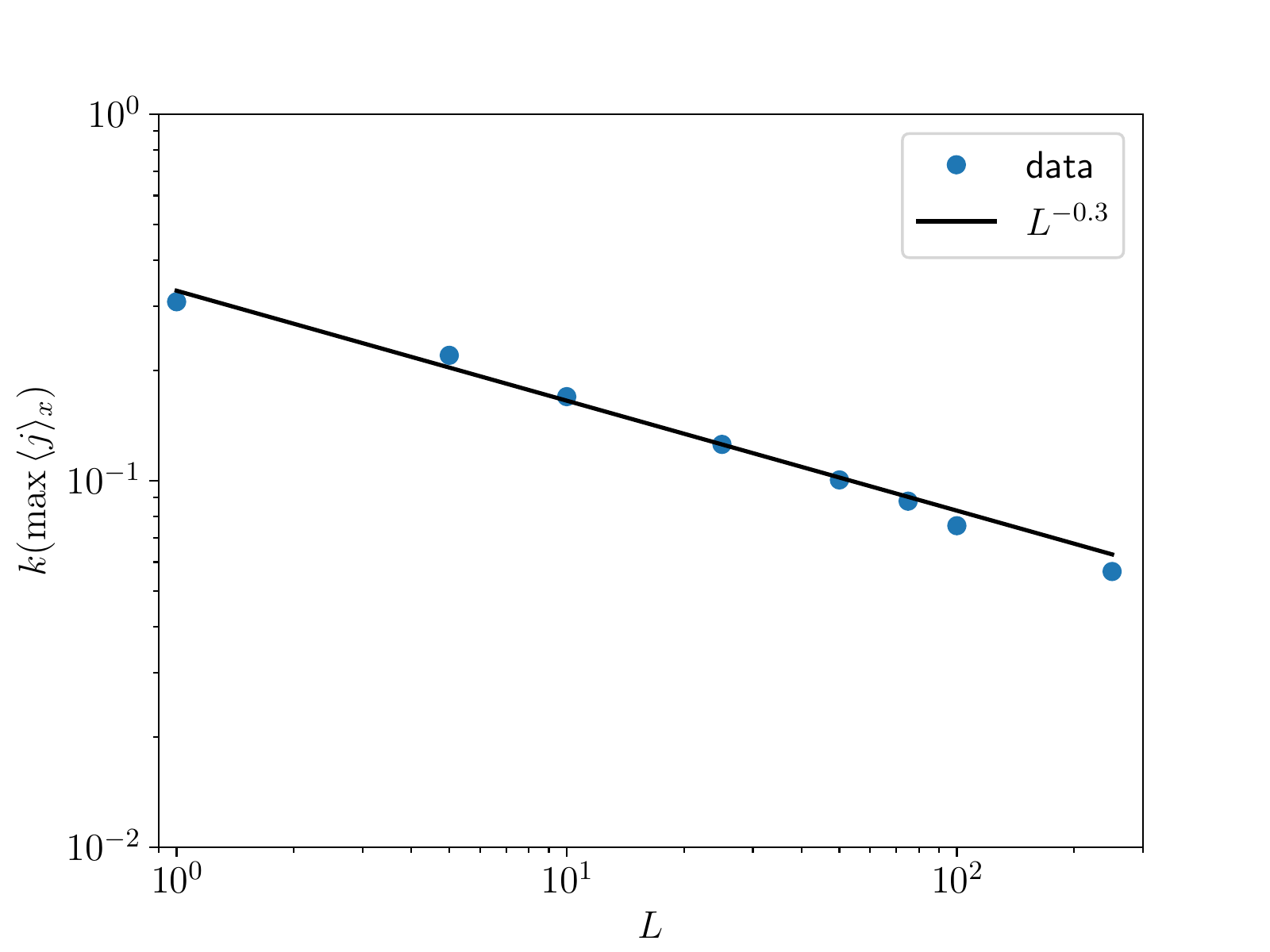}
  \caption{Scaling of $k({\rm max}(\langle j\rangle_x))$ (defined as the point of maximum in Fig.~\ref{fig:jk}) as function of $L$. The observed scaling is consistent with $L^{-0.3}$, which implies an anomalous exponent $\alpha=0.4$.
  \label{fig:kjmax}}
\end{figure}  
In particular, Fig. \ref{fig:jLk} shows how for these modes the energy current is in inverse 
proportionality with the chain length $L$ as one would expect from Fourier's law, i.e. \eqref{eqn:cond} 
when $\mathcal{K}$ does not depend on $L$. There shoud be, then, a critical value $k_c$ above which the 
evolution of the energy is diffusion-dominated and below which the predominant transport mechanism is the 
purely ballistic one. Considering that at equilibrium the transport term and the collision integral 
should balance, and using dimensional arguments, one can find that $k_c \approx L^{-3/10}$
\cite{Dematteis2020}. A scaling consistent with this estimation can be found, in our simulations, for the value of $k$ for which $\langle j\rangle_x$ is maximal, as reported in figure \ref{fig:kjmax}, suggesting that this criterion could be used as 
a proxi for the determination of $k_c(L)$, to distinguish between diffusive and ballistic modes.

\subsection{Ballistic and diffusive propagation}

\begin{figure*}
  \centering
  \begin{subfigure}{\columnwidth}
    \includegraphics[width = \columnwidth]{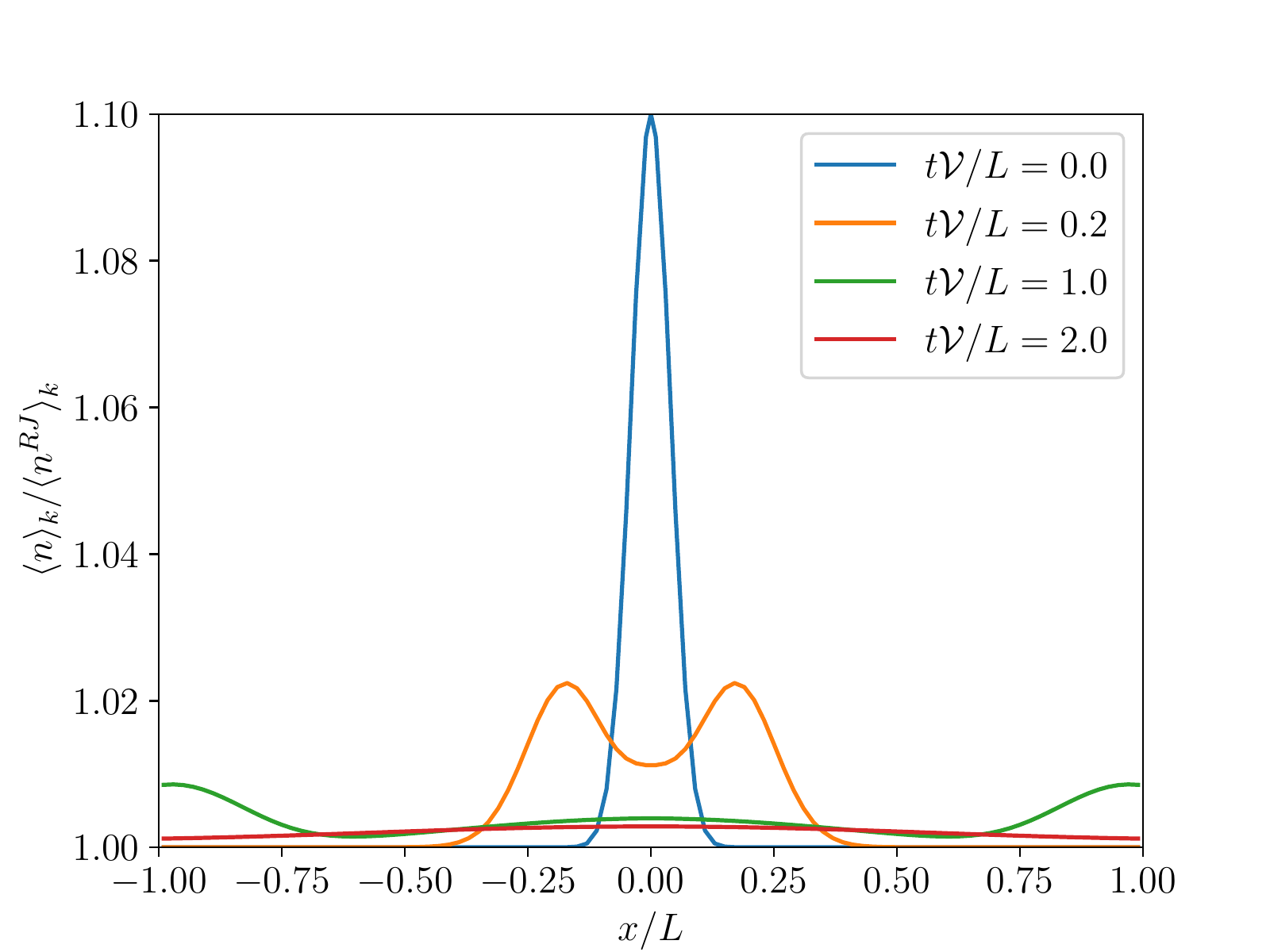}
    \caption{\label{fig:diffusion_n}}
  \end{subfigure}
  \begin{subfigure}{\columnwidth}
    \includegraphics[width = \columnwidth]{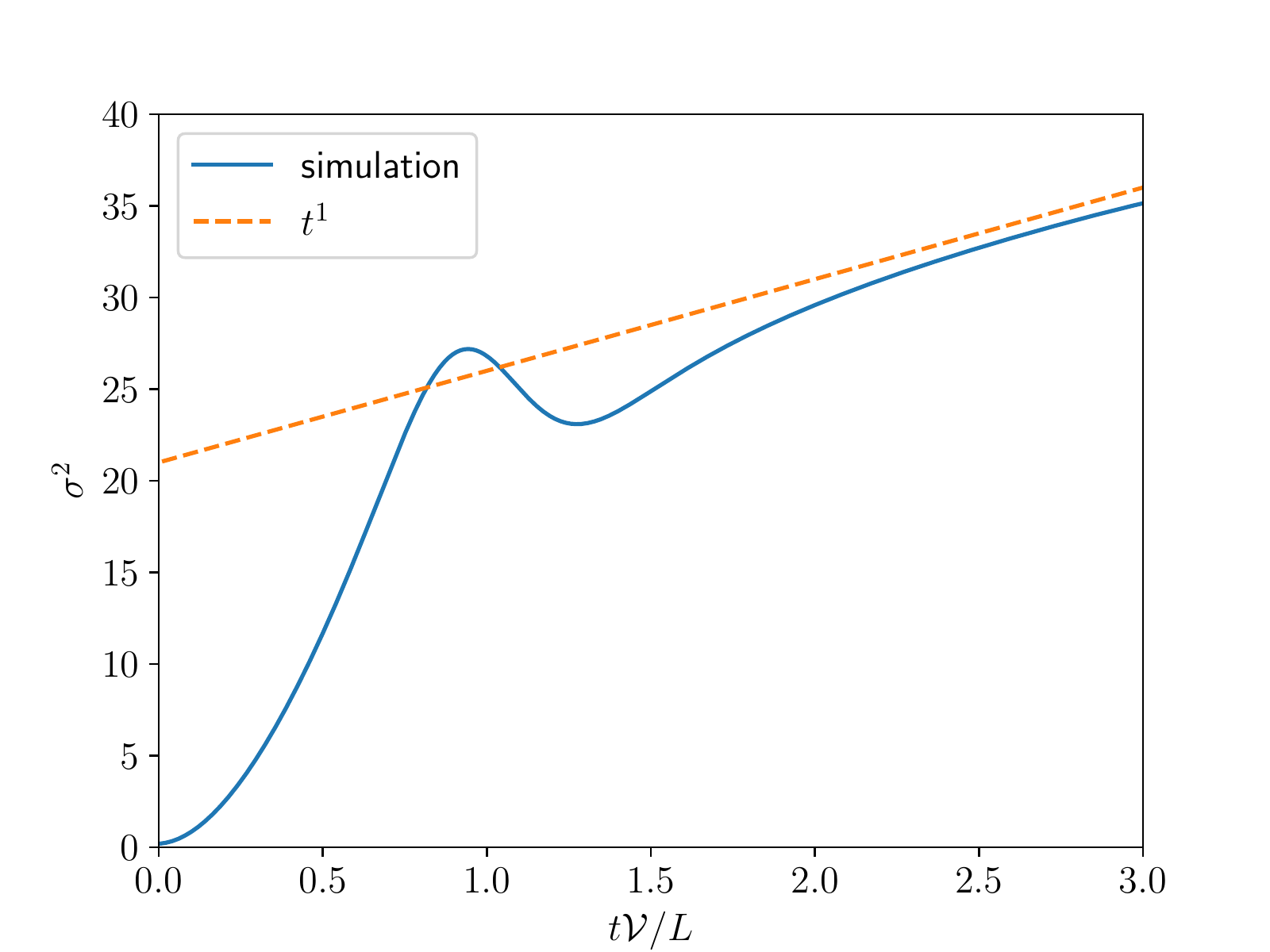}
    \caption{\label{fig:diffusion_sigma}}
  \end{subfigure}
  \caption{(a) Spatial distribution of $\langle n\rangle_k(x,t)$ (integrated in $k$) at different times. An initial spatially localized perturbation over a uniform background propagates as two ballistic peaks moving in opposite direction at constant velocity and a central decaying peak.  (b) Time   
    history of the variance of the temperature distribution. After the ballistic peaks exit the system, the broadening of the central peak (also known as the heat peak) shows agreement with a diffusive behavior. \label{fig:diffusion}}
\end{figure*}

\begin{figure*}
  \centering
  \begin{subfigure}{0.3\textwidth}
    \includegraphics[width = \textwidth]{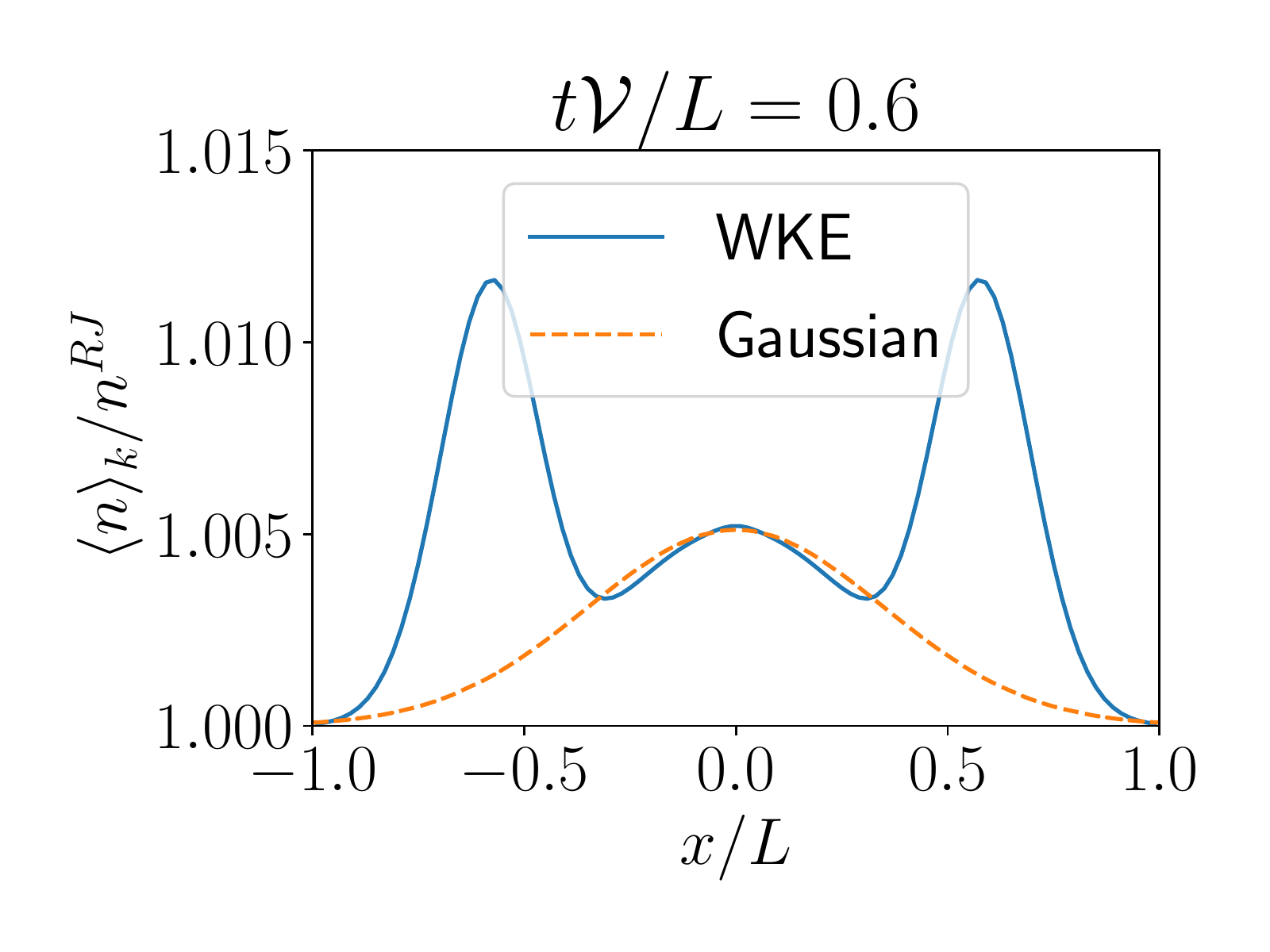}
    \caption{\label{fig:6a}}
  \end{subfigure}
  \begin{subfigure}{0.3\textwidth}
    \includegraphics[width = \textwidth]{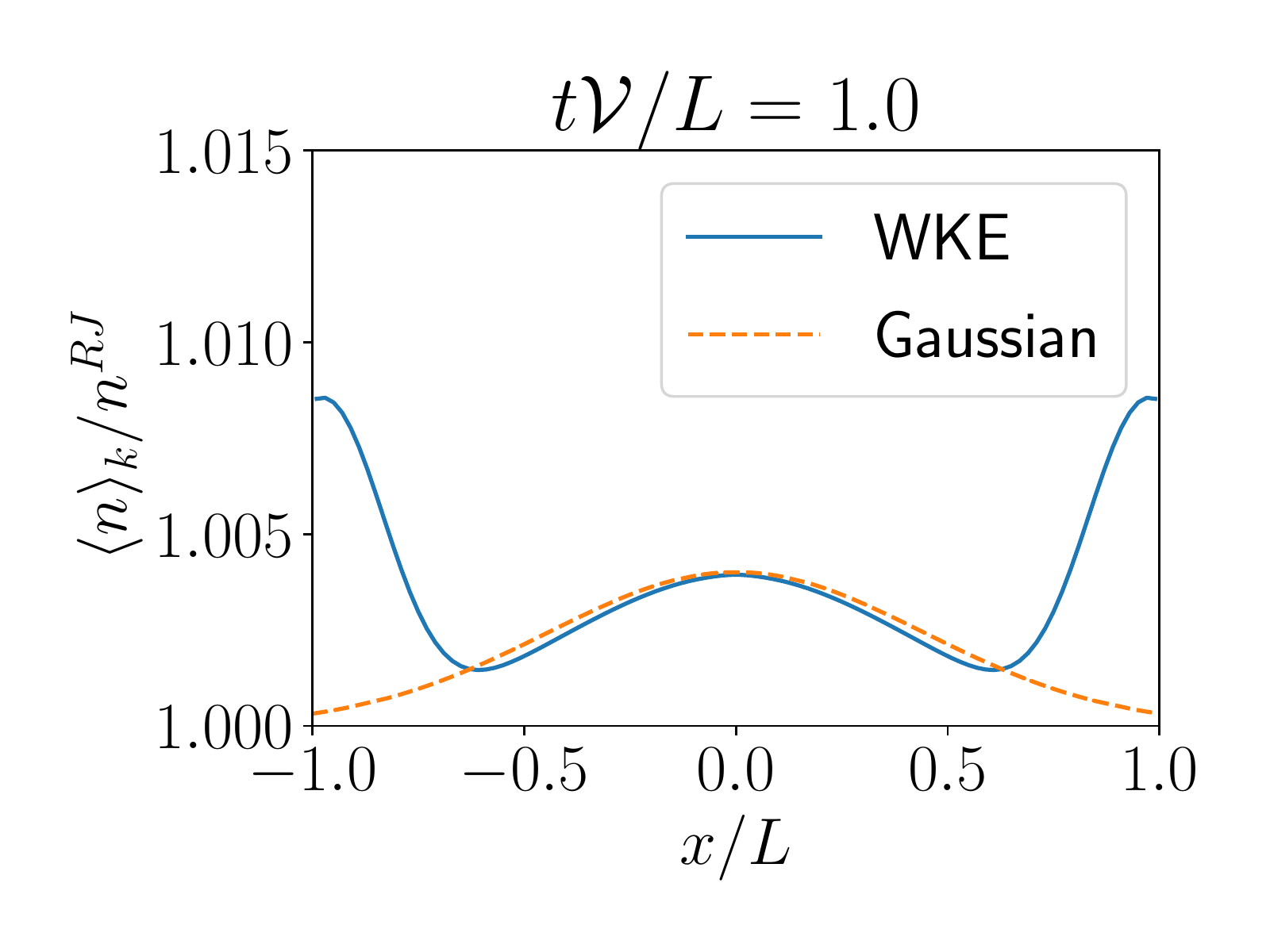}
    \caption{\label{fig:6b}}
  \end{subfigure}
  \begin{subfigure}{0.3\textwidth}
    \includegraphics[width = \textwidth]{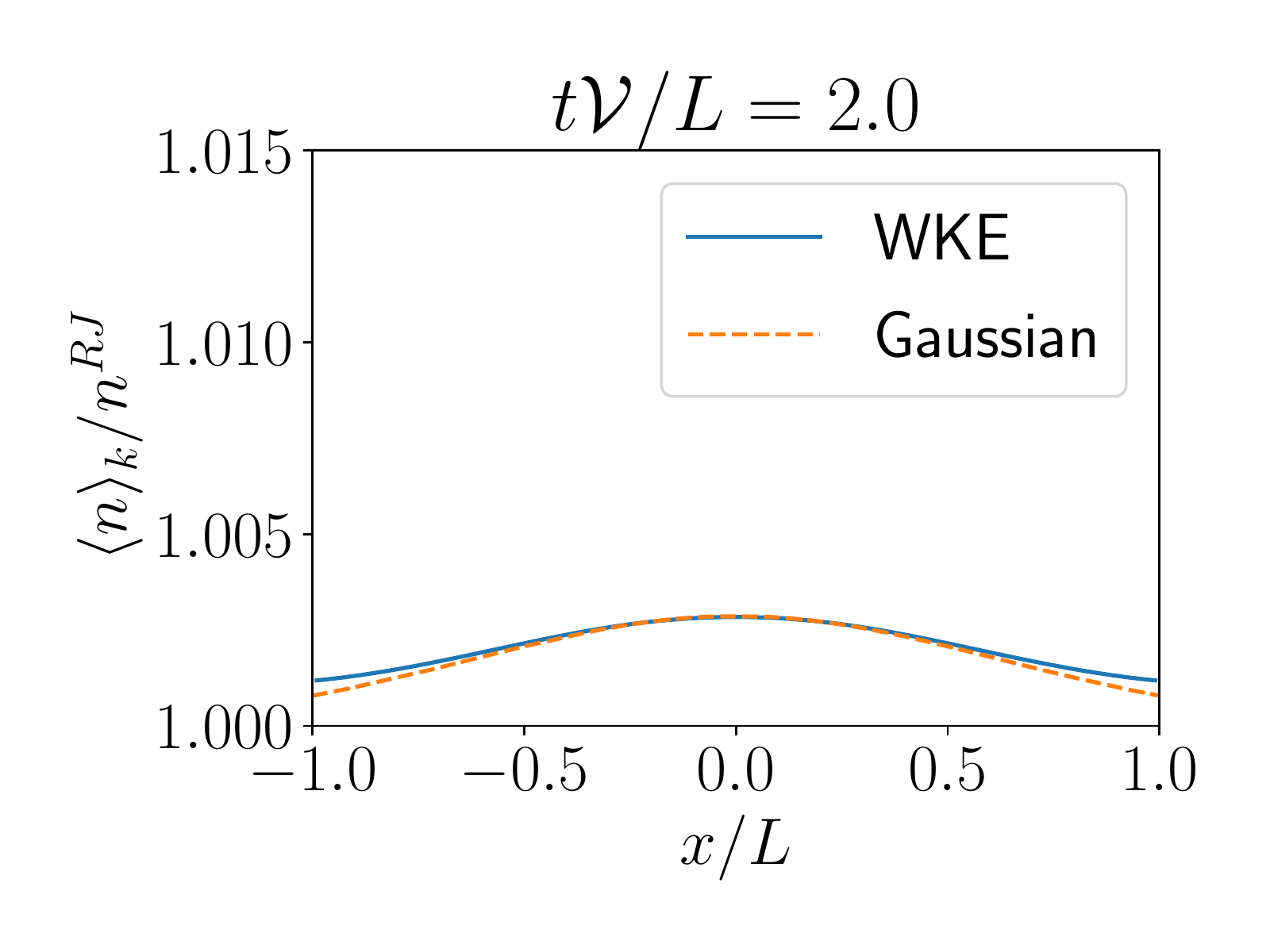}
    \caption{\label{fig:6c}}
  \end{subfigure}
  \caption{Heat peak evolution as predicted by WKE (\textcolor{blue}{--}), and pure diffusion 
    (\textcolor{red}{- -}).\label{fig:diffusion_comparison}}
\end{figure*}
\begin{figure*}
  \centering
  \includegraphics[width = \textwidth]{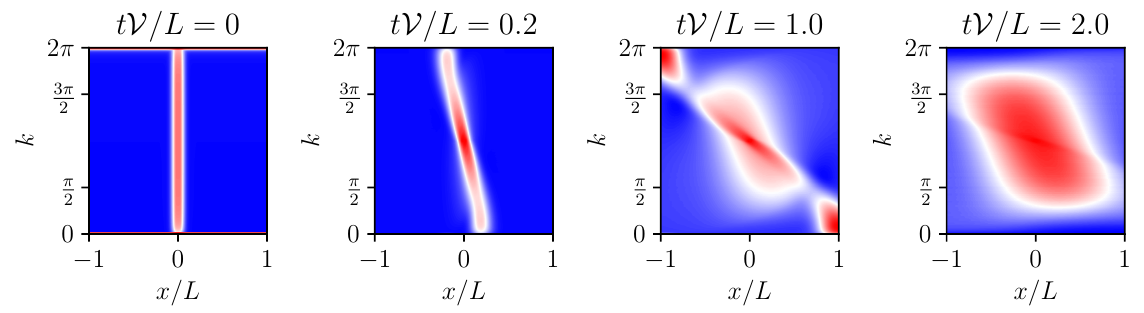}
  \caption{Color map of $e(x,k,t)=\omega_kn(x,k,t)$ at four non-dimensional time instants. The second sound emission is clearly seen in the central maps where perturbations at $k\simeq 0$ and $k\simeq 2\pi$ (the low modes) are detaching from the central diffusive peak involving the high modes. The interacting and diffusive character of these modes is evident from the fact that the shape of the central peak remains close to a rectangle during the evolution, tending to populate all modes $k$ with the same energy density at fixed position $x$. \label{fig:energy}}
\end{figure*}

In nonequilibrium statistical physics, the transport coefficients characterizing nonequilibrium steady 
states that are not too far from equilibrium can be computed in terms of space-time correlations in an 
equilibrium ensemble of realizations of the microscopic dynamics~\cite{kubo2012statistical,lepri2003thermal,aoki2006energy,lukkarinen2016kinetic,spohn2016fluctuating}, 
via the so-called Kubo integral. Likewise, for the mesoscopic model of~\eqref{eqn:wavkin}, let us now 
consider an initial background equilibrium state with constant $T = T_0$ and chemical potential $\mu$. 
Let us then consider an initial narrow bell-shaped perturbation $\delta T(x)$ in the center of the 
domain, such that $\delta T\ll T_0$. We initialize $n(x,k,t=0)$ with a RJ distribution, see~\eqref{eq:5}, 
with $T(x)=T_0+\delta T(x)$, {in a domain going from $-L$ to $L$}. 

The evolution of $ \langle n \rangle_k(x,t) = \int_0^{2\pi} n(x,k,t) dk$ is shown in Fig.~\ref{fig:diffusion}, where we considered a perturbation having an initial amplitude of $\delta T(x=0)/T_0 = 0.1$.
One can recognize the familiar behavior of a central 
peak ({\it heat peak}) and two traveling peaks ({\it acoustic peaks}) which correspond to the emission of the {\it second sound}.  This configuration has been studied 
 using the the microscopic dynamics~\cite{lepri2003thermal,lukkarinen2016kinetic} and the 
stochastic model known as {\it fluctuating hydrodynamics}~\cite{spohn2016fluctuating}.  
For $t>0$, two 
peaks separate and propagate in opposite directions with constant velocity about $\pm 1$; the central 
peak, instead, evolves diffusively in time (Fig.~\ref{fig:diffusion_n}). This is clearly visible by looking at 
the time evolution of the variance of the distribution, computed as follows:
\begin{equation}
  \sigma^2 = \frac{\int_{x,k} n(k,x,t)\omega(k) x^2 dx dk}{\int_{x,k} n(k,x,t)\omega(k) dx dk}.
\end{equation}
Indeed, as shown in Fig.~\ref{fig:diffusion_sigma}, after the acoustic peaks exit the domain (at 
about $t\mathcal{V}/L = 1$) and the central peak  is left alone, the variance starts to grow linearly in
time. The time evolution of the central peak follows 
regular diffusion, with diffusion coefficient given by half the slope of the asymptote on the right 
hand-side of Fig. \ref{fig:diffusion_sigma}. This particular result may sound controversial in the face 
of notable results advocating for a {\it heat peak} that follows {\it fractional diffusion} 
(of super-diffusive type). We address this further in the {\it Discussion} section.

We can therefore identify the second sound emission with the non-decaying transport of the ballistic modes, and the 
{\it heat peak} with the regular diffusion of the modes that thermalize locally. Further confirmation of 
this is found in Fig.~\ref{fig:diffusion_comparison}, where we plot the time 
evolution of the small initial Gaussian perturbation on homogeneous background, and we show
that the numerical simulation of the WKE follows closely a diffusive 
solution with diffusion coefficients
around $0.24$, for this condition. Thus, we can simply refer to the 
{\it heat} and the {\it acoustic} peaks as to the {\it diffusive} and the {\it ballistic} (or {\it second-sound}) peaks, 
respectively, without ambiguity. Fig.~\ref{fig:energy} shows the energy density $ e(x,k,t) $ at various times of the evolution of the perturbation. The low modes, with $k \approx 0$ and $k \approx 2\pi$, are the ones with the highest ballistic 
velocity. Hence, they will leave the domain in a timescale of the order of $L/\mathcal V$. For longer times, the higher modes start
to diffuse thanks to the collision integral and the distribution will start to follow a diffusive evolution. 

\section*{Discussion}

Our direct numerical simulation of the WKE shows that two phononic  states coexist in the $\beta$-FPUT chain. The first, involving the low modes, is equivalent to the emission of second sound; the second, involving the higher modes, is purely diffusive. This has been analyzed under the following different points of view.
\begin{itemize}
	\item The anomalous scaling of the energy  conductivity $\mathcal K\propto L^\alpha$, with $\alpha\simeq0.4$, is confirmed in Figs.~\ref{fig:KT}-\ref{fig:KL}. This is due to the scaling $ k_c(L) $ of the separation between the ballistic modes (low wavenumbers) and the diffusive modes (high wavenumbers), individually contributing towards $\alpha=1$ and $\alpha=0$, respectively. We find that $k_c$ varies consistently with $L$ by the scaling $k_c(L)\propto L^{-3/10}$, as shown in Fig.~\ref{fig:kjmax}.
	\item The spatial integral of the spectral energy current modal density $\langle j\rangle_x(k)$ is independent of $L$ for the low modes, as predicted for the purely harmonic chain \cite{rieder1967properties}, while it is proportional to $L^{-1}$ for the higher modes, in agreement with Fourier's law. This was shown in Fig.~\ref{fig:j}.
		\item A complementary way to analyze energy transport is to look at the evolution of a small localized perturbation of the thermal equilibrium condition. By doing that, we confirm the presence of two {\it acoustic} peaks shooting off in opposite directions and a central {\it heat} peak (figure~\ref{fig:diffusion_n}) evolving in agreement with standard Fourier diffusion, as confirmed in Figs. ~\ref{fig:diffusion_sigma} and \ref{fig:6a}-\ref{fig:6c}.
	\item Existence of the two types of heat transfer,
ballistic and diffusive, is cleanly demonstrated in
Figs.~\ref{fig:Tmap} and \ref{fig:energy}. In Fig.~\ref{fig:Tmap} the qualitative difference between
these two types is most evident near the stationary
nonequilibrium state,
where the horizontal separation between the two different regions gives an intuitive visualization of $k_c$. Finally, in Fig.~\ref{fig:energy} we see an $x-k$ representation of the evolution of a perturbed equilibrium state. 
	Again, the sharp separation wavenumber, $k_c$, can be observed by eye. Not surprisingly, we discover that the acoustic peaks are made exclusively of noninteracting ballistic modes with low wavenumber, while the heat peak is made exclusively of modes with high wavenumber. 
\end{itemize}

Although the separation of scales at $k_c$ observed in the $ x-k $ plots is slightly smeared, the two-state ballistic-diffusive picture analyzed above under these four different angles is robust and does not include super-diffusive propagation with fractionary exponents. A fractionary diffusion equation with space derivative of order $8/5$ instead of $2$, is an alternative explanation compatible with the conductivity scaling $\mathcal \mathcal{K} \propto L^{2/5}$, rigorously derived in \cite{mellet2015anomalous}. However, this scaling would predict the propagation of a single peak that does not find correspondence in our observations.
The apparent disagreement may have different origins. For example, the assumption of having only two conserved quantities (energy and action) made in \cite{mellet2015anomalous} is partially violated by the ballistic modes, which effectively preserve momentum and whose propagation is not a part of the heat peak. Although we have assessed overall compatibility of the heat peak evolution with a diffusive behavior, providing a definitive study of this issue is not the aim of the current manuscript. 
In close analogy with {\it second sound} propagation in superfluids~\cite{peshkov2013second}, our results show that, if the ballistic phonons are recognized as noninteracting traveling waves~\cite{kuzkin2020ballistic,Dematteis2020,kuzkin2021unsteady}, the two-state ballistic-diffusive picture is compatible with the main observable aspects of energy transport.

Finally, it is worth noticing that second sound in dielectric solids was predicted a long time ago~\cite{chester1963second,prohofsky1964second,chandrasekharaiah1986thermoelasticity}, and later observed for instance in solid He$^3$ and He$^4$ below $4$ K and in NaF below $20$ K, at extremely low temperature. In a dielectric crystal, second sound can be observed when Umklapp resonances are very small, and by lowering the temperature enough, the scattering level is reduced to a point where noninteracting wavelike transport becomes visible on macroscopic scales. It is now well-known that reducing the dimensionality of the material is another way to reduce drastically the number of interactions, in part explaining why it was recently possible to observe second-sound propagation in 2D graphite at temperatures above $100$ K~\cite{huberman2019observation,chen2021non}. Our results further indicate that reducing the dimensionalty to (quasi-)1D structures such as nanotubes may give a hope to finally observe second-sound propagation at room temperature. 

\bigskip\bigskip

\section*{Materials and Methods}
\subsection*{Integration on the resonant manifold}
 In \eqref{eqn:coll}, the equality coming from the momenta Dirac delta has to be interpreted $\mod I, I=[0,2\pi)$, to include possible Umklap resonances. The resonant manifold is the subset of $I\times I\times I\times I$ satisfying at the same time the two conditions
\begin{equation}\label{eq:3m}
\omega_k+\omega_{k_1}-\omega_{k_2}-\omega_{k_3} = 0\,,\qquad k_3 = (k+k_1-k_2)\mod I\,.
\end{equation}
The constraint imposed by integration on the resonant manifold reduces the triple integral of \eqref{eqn:coll} to a one-dimensional integral.
Here, we briefly report some important rigorous results from Ref.~\cite{lukkarinen2008anomalous} (see also~\cite{lukkarinen2016kinetic,pereverzev2003fermi}).
The solutions of the collisional constraints \eqref{eq:3m} are of three types:
\begin{itemize}
\item $k_1 = k_3$, $k_2=k_4$\,;
\item $k_1 = k_4$, $k_2=k_3$\,;
\item $k_2=h(k_1,k_3) \mod I$, where
\begin{equation}
	h(x,y) = \frac{y-x}{2} + 2 \arcsin\left( \tan{\frac{|y-x|}{4}} \cos \frac{y+x}{4} \right)\,.
\end{equation}
\end{itemize}
The first two types (perturbative solutions) are trivial resonances that contribute to nonlinear frequency shift and broadening~\cite{lvov2018double}. The third type of solutions (non-perturbative) represents non-trivial resonances that are responsible for irreversible spectral transfers.
By integrating analytically in $k_3$, the collision integral of \eqref{eqn:coll} can be written as
\begin{equation}\label{eq:5m}
	\mathcal I_k = \int_{0}^{2\pi} dk_1dk_2\; g(k,k_1,k_2) \delta(\Omega(k,k_1,k_2))\,,\\
\end{equation}
with
\begin{equation}\label{eq:6m}
\begin{aligned}
	g(k,k_1,k_2) &= |T_{k,k_1,k_2,k+k_1-k_2}|^2 n_kn_{k_1}n_{k_2}n_{k+k_1-k_2}\\
				& \;\;\;\times\left(\frac{1}{n_k}+\frac{1}{n_{k_1}}-\frac{1}{n_{k_2}}-\frac{1}{n_{k+k_1-k_2}} \right)\,,\\
\Omega(k,k_1,k_2) &= \omega_k+\omega_{k_1}-\omega_{k_2}-\omega_{k+k_1-k_2}\,.
\end{aligned}
\end{equation}
In order to integrate out the frequency delta, we exploit the following property of the Dirac delta function:
\begin{equation}\label{eq:7m}
	\int dx \;G(x)\delta(f(x)) = \int dx \;G(x) \sum_i \frac{\delta(x-x_i^\star)}{|f'(x_i^\star)|}\,,
\end{equation}
where $x_i^\star$ are all the zeros of $f$. In \eqref{eq:5m}, integrating in the variable $k_1$, we know that all of the zeros of $\Omega = 2[\sin(k/2) + \sin(k_1/2) - \sin(k_2/2) - \sin(|k+k_1-k_2|/2)]$ are of one of the three types above. The trivial solutions give $\Omega'(k_1)=0$ identically, which implies singlular denominators. Though, as discussed in Ref.~\cite{lukkarinen2008anomalous} these terms come in pairs of opposite sign (this can be seen easily looking at the symmetries of the integrand of~\eqref{eqn:coll}), which cancel each other and do not contribute. Therefore, the non-vanishing contributions come from the non-trivial resonances, and we obtain
\begin{equation}\label{eq:8m}
\begin{aligned}
\mathcal I_k&= \int_0^{2\pi} dk_2 \int_0^{2\pi}dk_1 \; g(k,k_1,k_2)\frac{\delta(k_1-h(k,k_2))}{|\partial_{k_1}\Omega(k,h(k,k_2),k_2)|}\\
&= \int_0^{2\pi} dk_2 \; \frac{g(k,h(k,k_2),k_2)}{\sqrt{\left(\cos\tfrac{k}{2}+\cos\tfrac{k_2}{2}\right)^2 + 4 \sin\tfrac{k}{2}\sin\tfrac{k_2}{2}}}\,.
\end{aligned}
\end{equation}

\subsection*{Numerical details}
\eqref{eqn:wavkin} is solved by finite difference approximation in time and space, using the expression
 of $\mathcal I_k$ given in \eqref{eq:8m}. 
In all simulations we used 100 grid points in $x$ and 1001 points in $k$; the chemical potential is set
to $\mu = 0.05$. The adopted discretization guarantees the conservation of energy and wave action. In {\it Case A} of the {\it Results} we used $T_1=0.4$, $T_2=0.2$. In {\it Case B} of the {\it Results} we used $T_0=0.3$, $\delta T(x=0)/T_0=0.1$.

\section*{Acknowledgements}
M.O.  was supported by the ``Departments of Excellence 2018-2022'' Grant awarded by the Italian Ministry of Education, University and Research (MIUR) (L.232/2016). GD and YL gratefully acknowledge funding from ONR grant N00014-17-1-2852. YL also acknowledges support from NSF DMS award 2009418. M.O. was supported by Simons Collaboration on
Wave Turbulence, Grant No. 617006. We thank Gregory Falkovich for pointing out to us the analogy with second-sound propagation. Lamberto Rondoni is also acknowledged for fruitful discussions during the early stages of the work.


\bibliography{biblio}

\end{document}